\documentclass{PoS}

\title{ Progress on the Microscopic Spectrum of the Dirac Operator for QCD with Wilson Fermions}

\ShortTitle{Progress on Wilson Dirac Spectrum}

\author{K. Splittorff\\
       Discovery Center, The Niels Bohr Institute \\
       University of Copenhagen, \\
       Blegdamsvej 17, DK-2100 Copenhagen, Denmark\\
       E-mail: \email{split@nbi.dk}}
\author{\speaker{J.J.M. Verbaarschot}\\
       Stony Brook University\\
       Department of Physics and Astronomy\\
       Stony Brook, NY 11794-3800, USA\\
       E-mail: \email{jacobus.verbaarschot@stonybrook.edu}}

\abstract{
Starting from the chiral Lagrangian for Wilson fermions at nonzero
lattice spacing we have obtained  compact expressions for 
all spectral correlation functions of the Hermitian Wilson Dirac
operator in the $\epsilon$-domain of QCD with dynamical quarks.
We have also obtained the distribution of the chiralities
over the real eigenvalues of the Wilson Dirac operator for any 
number of flavors.
All results have been derived for a fixed index of the Dirac 
operator. 
An important effect of dynamical quarks 
is that they completely suppress
the inverse square root singularity in the spectral density of the
Hermitian Wilson Dirac operator.
The analytical results are given in terms of an integral
over a diffusion kernel for which the square of the lattice spacing 
plays the role of time.
 This approach greatly simplifies the
expressions which we here reduce to the evaluation of
two-dimensional integrals. 
}

\FullConference{ The XXIX International Symposium on Lattice Field Theory - Lattice 2011\\
July 10-16, 2011\\
Squaw Valley, Lake Tahoe, California}

\begin{document}

\newcommand{\colbs}{}
\newcommand{\be}{\begin{eqnarray}}
\newcommand{\ee}{\end{eqnarray}}
\newcommand\del{\partial}
\newcommand\nn{\nonumber}
\newcommand{\Tr}{{\rm Tr}}
\newcommand{\Str}{{\rm STr}}
\newcommand{\mat}{\left ( \begin{array}{cc}}
\newcommand{\emat}{\end{array} \right )}
\newcommand{\vect}{\left ( \begin{array}{c}}
\newcommand{\evect}{\end{array} \right )}
\newcommand{\tr}{{\rm Tr}}
\newcommand{\hm}{\hat m}
\newcommand{\ha}{\hat a}
\newcommand{\hz}{\hat z}
\newcommand{\hx}{\hat x}
\newcommand{\tm}{\tilde{m}}
\newcommand{\ta}{\tilde{a}}
\newcommand{\tz}{\tilde{z}}
\newcommand{\tx}{\tilde{x}}
 \newcommand{\bitem}{\begin{itemize}}
\newcommand{\eitem}{\end{itemize}}
 \newcommand{\bmini}{\begin{minipage}}
\newcommand{\emini}{\end{minipage}}
 \newcommand{\dl}{$}
\newcommand{\dr}{$}

\section{Introduction and Motivation}

The low-lying spectrum of the continuum Dirac operator has been
described successfully in terms of chiral Lagrangians or Random 
Matrix Theory \cite{SV,DOTV}. 
The reason for this success is that the smallest eigenvalues
are of order $1/V$, which is deep inside the $\epsilon$-domain. In this 
domain
the static part of the chiral Lagrangian, which is equivalent to
Random Matrix Theory,  
factorizes from the partition
function, and determines
properties of the low-lying Dirac eigenvalues.
In this lecture we discuss the extension of
these results  to nonzero lattice spacing.
 We can distinguish two different types of discretization
effects: those that do not affect the symmetries of
the Dirac operator and those that do affect its symmetries. In the first
case, discretization effects can be absorbed into a redefinition of
the low-energy constants. In the second case, the discretization errors
take us outside the universality class which requires additional
terms in the chiral Lagrangian.
The two main discretization schemes are staggered fermions and Wilson
fermions and both break the symmetries of the continuum Dirac operator.
In this talk we will discuss recent results 
\cite{DSV,ADSV,ADSVNf1,SV11} for the  
the Wilson Dirac operator including its dependence on the topological
index and the number of flavors.  
The confirmation of these results by lattice simulations 
\cite{heller,deuzeman} will be discussed elsewhere in these 
Proceedings. For recent progress on the staggered Dirac operator 
see \cite{Osborn}.

We will consider discretization effects of the Wilson
Dirac operator
\be
D_W = \frac 12 \gamma_\mu( \nabla_\mu +\nabla_\mu^*) - \frac 12 a 
 \nabla_\mu ^*\nabla_\mu.
\ee 
Because of its $\gamma_5$ Hermiticity it is advantageous to analyze 
the spectrum of the Hermitian Dirac
operator  \cite{Luscher,Golterman,Sharpe,NS}
 \be D_5 \equiv \gamma_5(D_W+m) =D_5^\dagger .\ee
Since $D_W$ is nonhermitian, its eigenvalues are complex, but because
of the $\gamma_5$ Hermiticity, they occur in complex conjugate pairs or
are real. Therefore the complex eigenvalues can
collide with the real axis and turn into a pair of real eigenvalues. This is
different from QCD at nonzero chemical potential \cite{AOSV} where the nonzero 
Dirac eigenvalues are complex and occur in pairs of opposite sign. 
Then both their real and  imaginary parts have 
to vanish for a pair to collide with the real axis. 
 
In this talk we report on results for discretization effects on spectra
of the Wilson Dirac operator of QCD with $dynamical$ quarks. A direct evaluation 
of the generating function is complicated, but using a
 diffusion method \cite{Guhr} we obtain compact expressions
for correlation functions of Dirac eigenvalues which we here further
simplified by exploiting an underlying Pfaffian structure \cite{Kieburg}.

We start from a chiral Lagrangian, but our results can also be derived
from the corresponding chiral Random Matrix Theory \cite{DSV,ADSV}. Recently, the joint 
eigenvalue distribution of both the Wilson Dirac operator \cite{mario} 
and the hermitian Wilson Dirac operator \cite{nagao} have been obtained from RMT.

\section{Spectrum of the Dirac Operator}

In order to access the spectrum of $D_5$, we introduce an axial mass $z$
\be
\det(D_w +m + \gamma_5 z) = \det(\gamma_5(D_W+m) +z).
\ee
The low-energy limit of the corresponding partition function is given by
a chiral Lagrangian that up to low energy constants is uniquely determined
by symmetries \cite{SharpeSingleton,RS,BRS}
\be
{\cal L}(m,z;a)= \frac 12 m\Sigma \Tr (U+ U^\dagger) 
+\frac 12 z\Sigma \Tr (U -U^\dagger)  -a^2 W_8 \Tr( U^2 +U^{-2}). 
\ee
In the microscopic domain, where the combinations
$mV$, $zV$ and $ a^2 V$ are kept fixed in the thermodynamic limit, the
$m$, $ z$ and $a$-dependence of the partition function resides in its
zero momentum part that factorizes from the nonzero momentum part \cite{DSV}
\be
Z^\nu_{N_f}(m,z;a)=\int_{U \in U(N_f)} dU {\det}^\nu U e^{\frac 12 mV\Sigma \Tr (U+ U^\dagger) 
+\frac 12 zV\Sigma \Tr (U -U^\dagger)  -a^2 V W_8 \Tr( U^2 +U^{-2})}. 
\label{Z}
\ee
To this order the chiral Lagrangian also contains
$
-a^2 W_6 [{\rm Tr} (U+U^\dagger)]^2
-a^2 W_7 [{\rm Tr} (U-U^\dagger)]^2\nn
$
but they can be added to the mass term after linearizing the squares at
the expense of a Gaussian integral and will not be considered below.

Since small Dirac eigenvalues are very sensitive to the index, $\nu$, of the Dirac operator 
we will work at fixed $\nu$. This index is determined from the spectral flow of the
eigenvalues of $D_5(m)$ \cite{Itoh,Heller} (see Fig. 1).
At the crossing point, $m_c$, with the real axis  we have
(for small $a$, the physical part of the lattice Dirac spectrum can 
be separated from the unphysical part)
\be
\gamma_5(D_W+m_c)\phi =0 \quad \Longrightarrow \quad D_W \phi = -m_c \phi.
\ee
Note that multiple crossings may occur. Therefore the number of real 
eigenvalues of $ D_w$  is at least equal to its index.
The total number of flow lines with a net flow across the real axis
 is a topological
invariant of the Dirac operator.
\begin{figure}
\centerline{\includegraphics[height=2.1cm,]{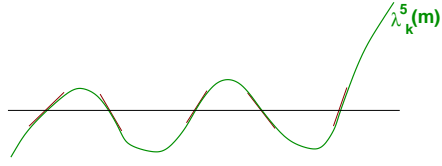}}
\caption{Spectral flow of eigenvalues of $D_5$ versus the quark mass $m$.}
\end{figure}
It can be written as a sum  over eigenvalues of  $ D_5 $~\cite{Itoh}
\be
\nu = \sum_{m_k, \lambda^5(m_k) = 0} {\rm sign} \left ( \frac {d\lambda^5(m)}{dm}
\right )_{m=m_k}=
\sum_k {\rm sign}(\langle \phi_k|\gamma_5 |\phi_k \rangle).
\label{nu-flow} 
\ee
The $\theta$-dependence of the QCD partition function is given by
\be
Z_{N_f}(m,z;a,\theta)= \sum_\nu e^{i\nu\theta} Z^\nu_{N_f}(m,z;a).
\label{zth}
\ee
In the continuum limit,  because of the anomaly, the factor 
$\exp(i\nu\theta)$ can be written as the Jacobian
of an axial transformation on the quark fields. At nonzero $a$ 
the natural extension is (See  \cite{arXiv:0811.0409}.)
\be
m \to e^{i\theta \gamma_5/N_f} m, \qquad
a \to e^{i\theta \gamma_5/N_f} a.
\label{rescaling}
\ee
As is the case in the continuum limit,
the partition function at fixed $\nu$ is defined by inverting 
the decomposition (\ref{zth}), which also can be applied
to the chiral Lagrangian
\be
Z^\nu_{N_f}(m,z;a) =\frac 1{2\pi} \int d\theta e^{i\nu\theta}Z_{N_f}(m,z;a,\theta),
\ee
where the $\theta$-dependence of the chiral Lagrangian is induced by 
the transformation (\ref{rescaling}).

The generating functions for the Wilson Dirac spectrum are given by
\cite{DSV}
\be
Z^\nu_{N_f+p|p}(m,z_k, z'_k; a) = \left \langle {\det}^{N_f}( D_W + m)\prod_{k=1}^p
\frac{\det(D_W +m+ \gamma_5 z_k) }{\det(D_W +m +\gamma_5 z'_k)}
\right \rangle_\nu.\nn
\ee
The resolvent $( p = 1 )$ and the spectral density are given by
\be
G^\nu_{N_f+1|1}(z,m;a) = \left .\lim_{z' \to z} \frac d{dz} Z^\nu_{N_f+1|1}(m,z,z';a) \right |_{z'=z},
\qquad
\rho_5^\nu(\lambda^5,m;a) = \frac 1\pi {\rm Im} \,G^\nu(z,m;a)\big|_{z=\lambda^5}.
\nn
\ee
For $z \ll \Lambda_{\rm QCD} $  the $ z$-dependence of the generating
function is given by a chiral
Lagrangian that is uniquely determined by symmetries that are compatible
with the convergence of the bosonic integrals. In the microscopic
domain the generating function reduces to the graded integral \cite{DSV,DOTV}
\be 
Z_{N_f+1|1}^\nu(m,z,a)=\frac 1{Z_{N_f}^\nu(m;a)}
\int_{U \in Gl(N_f+1|1)} \hspace{-8mm} dU {\rm Sdet}^\nu(U)e^{ {\colbs i}\frac 12 mV\Sigma \Tr (U- U^\dagger) 
+{\colbs i}\frac 12 V\Sigma \Tr (\zeta U +U^\dagger\zeta)  -
{\colbs i^2}a^2 V W_8 \Tr( U^2 +U^{-2})}, \nn\\[-0.4cm]
\label{zgen}
\ee
where, $\zeta_3 = {\rm diag} (0,\cdots, 0,z,z')$.
 The transformation $ U \to iU$ is required to get  convergent 
integrals in the noncompact sector for $ W_8 > 0$ (for which the Aoki phase 
\cite{Aokiclassic} can occur).

\begin{figure}
\hspace*{0.1cm}
\bmini{6cm}
\centerline{\includegraphics[height=4cm,angle=0,clip]{ga5DW-Nf01and2-nu0.eps}
}\emini
\hspace*{1cm}
\bmini{6cm}
\centerline{
\includegraphics[height=7.0cm, angle=-90,viewport= 0 0 500 700,clip]{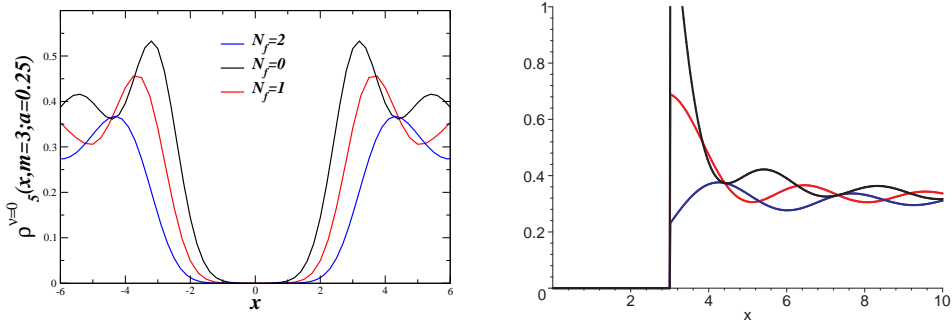}
}\emini
\caption {The spectral density of $\gamma_5(D_W +m)$
for $ m =3 $, $ \nu = 0$ and $a =0$ (right) and  $a=0.25$ (left) 
for $N_f =0$, $N_f =1$ $N_f =2$.}
\label{fig1}
\end{figure}

A direct evaluation of the integrals over the supergroup becomes
very laborious for $ N_f \ge 1$ but they still can be worked out for one flavor 
\cite{ADSVNf1,arXiv:1110.5744}. However, for
degenerate quark masses the expressions can be simplified
by starting from the identity 
\be
e^{ a^2 {\rm Trg}(U^2+{U^\dagger}^2)}&=&
e^{-2N_f a^2+ a^2 {\rm Trg}(U+U^\dagger)^2}
=e^{-2N_f a^2}\int d\sigma e^{ {\rm Trg} \sigma^2 /16 a^2+ \frac i2 {\rm Trg}\sigma (U+{U^\dagger})},
\nn  
\ee
 where $ \sigma $ is an $ (N_f+p|p)$ graded ``Hermitian'' matrix.
  In a diagonal representation of $\sigma$  the partition function
can be expressed in terms of an integral over shifted quark masses.
The result for the resolvent  is given by  \cite{SV11}
\be
G_{N_f+1|1}^\nu(m,z; a)
& =& 
\frac 1{Z_{N_f}^\nu(m;a)}
 \int dS J(S)
e^{(1/16a^2){\rm Trg}(S-z)^2} {\rm Sdet}^\nu(S-m)
Z_{N_f+1|1}^\nu\big(\{ \sqrt{{m}^2-S_k^2} \} 
; a= 0 \big), \nn\\[-0.3cm]
\label{G_N_f+p|p}
\ee
where $ J(S) $ is the Jacobian of transforming $\sigma$ to the
diagonal representation $ S $.


\subsection{Effect of Light Flavors}

 For $a= 0$ the spectral density of $ \gamma_5(D_W +m)$ is obtained by a simple
variable transformation of the spectral density $\rho^\nu(x)$ of $D_W$:
$\rho_5^\nu(x) =  x/{\sqrt{x^2-m^2}} \rho^\nu(\sqrt{x^2-m^2})\theta(|x|-|m|)$
(See Fig. \ref{fig1} (right)). For $a\ne0$ the result for the spectral density
follows from the resolvent (\ref{G_N_f+p|p}) (see Fig. \ref{fig1} (left)).
Although, for $N_f =1$ and $N_f =2$ the integrals in (\ref{G_N_f+p|p})
can still be evaluated directly, it becomes computationally expensive. However,
because of an underlying Pfaffian structure \cite{Kieburg}, 
the integrals can be
written as the sum of products of two and one dimensional integrals.

 \begin{figure}
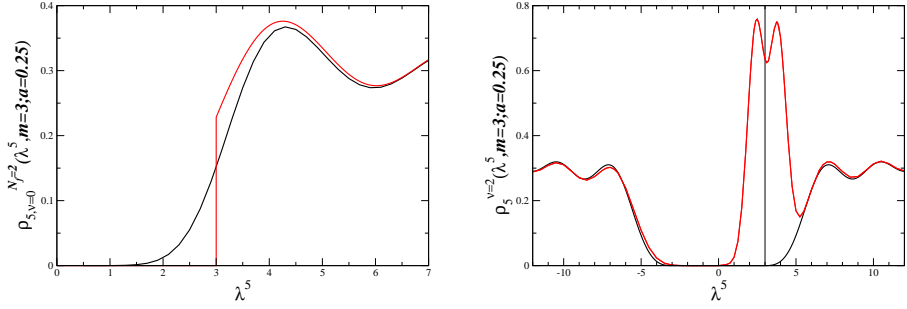

\centerline{ \includegraphics[height=4cm,angle=0,clip]{rho5_Nf2_a0.25anda0PRL_m3_nu0.eps}
\hspace*{0.5cm}
\includegraphics[height=4cm,angle=0,clip]{rho5_Nf2_a0.25_m3_nu2.eps}}
\caption
{Spectral density for two flavors with quark mass $ mV\Sigma = 3$
for $\nu = 0 $ (left) and $ \nu = 2 $ (right). We show the result for
  $ a=0 $  (black) and $ aW_8 \sqrt V =0.25$  (red).
\label{fig2}}
\end{figure}
The three 
dimensional 
integral
for one flavor resolvent (\ref{G_N_f+p|p}) can be rewritten as 
\be
G^\nu_{2|1}(z,m;a) = G_{1|1}^\nu(z,z,m;a) + G_{1|1}^\nu(0,z,m;a)
\frac{Z_1^\nu(z,m;a)}{Z_1^\nu(0,m;a)}
-Z_2^\nu(0,z,m;a) \frac{Z^\nu_{0|1}(z,m;a)}{Z_1^\nu(0,m;a)},\hspace*{0.5cm}
\ee
where the first term is the quenched resolvent $G_{1|1}^\nu(z,z,m;a)$
with
\be
G_{1|1}^\nu(z_1,z_2,m;a) &=& \frac {-1}{16\pi a^2} \int \frac{ds dt}{t-is}
e^{-[(s+iz_1)^2+(t-z_2)^2]/16a^2} \frac{(m-is)^\nu}{(m-t)^\nu}
\tilde Z_{1|1}^\nu((m^2+s^2)^{1/2},(m^2-t^2)^{1/2}).\hspace*{0.8cm}\nn
\ee
The various partition functions that enter in the expression are given by
\be
Z_{0|1}^\nu(z,m;a )&=& \frac 1{\sqrt{16\pi a^2}}\int dt \ e^{-(t-z)^2/16a^2}
\frac{(m^2-t^2)^{\nu/2}}{(t-m)^\nu}K_\nu((m^2-t^2)^{1/2}),\nn\\
Z_{1}^\nu(z,m;a )&=& \frac 1{\sqrt{16\pi a^2}}\int ds \ e^{-(s+iz)^2/16a^2}
\frac{(is-m)^\nu}{(m^2+s^2)^{\nu/2}}I_\nu((m^2+s^2)^{1/2}),\nn\\
Z_2^\nu(z_1,z_2,m;a) &=&  \frac 1{16\pi a^2}\int ds_1 d s_2 (is_1-is_2)
e^{[(is_1-z_1)^2+(is_2-z_2)^2]/16a^2} (is_1-m)^\nu (is_2-m)^\nu\ \\
&&\times
\tilde{Z}_2^\nu((m^2+s_1^2)^{1/2},(m^2+s_2^2)^{1/2}) ,\nn\\
\tilde Z^\nu_{1|1}(x,y) &=& \frac{y^\nu}{x^\nu}[yK_{\nu+1}(y)I_\nu(x)+
xK_\nu(y)I_{\nu+1}(x)], \quad 
\tilde Z_2^\nu(x, y) = \frac{yI_{\nu+1}(y)I_\nu(x)-xI_{\nu+1}(x)I_\nu(y)}
{x^\nu y^\nu(y^2-x^2)}.\nn
\ee
Likewise the four-dimensional integral (\ref{G_N_f+p|p}) for the $N_f=2$ resolvent    
be re-expressed as 
\be
&   & G_{3|1}^\nu(z,m;a) \\
& = & G_{1|1}^\nu(z,z,m;a)- \left .\frac {id G_{1|1}(z_1,z,m;a)}{dz_1} \right |_{z_1=0} \frac {Z_2^\nu(0,z,m;a)}{Z_2^\nu(m;a)}
+i\frac{G_{1|1}^\nu(0,z,m;a)}{{Z_2^\nu(m;a)}} 
\left . \frac {d Z_2^\nu(z_1,z,m;a)}{dz_1} \right |_{z_1= 0} \hspace*{-0.7cm},\hspace*{0.3cm} \nn
\ee
where
\be
Z_2^\nu(m;a) &=&  \frac i{\pi(16a^2)^2}\int ds_1 d s_2 (is_1-is_2)^2
e^{[(is_1-z_1)^2+(is_2-z_2)^2]/16a^2} (is_1-m)^\nu (is_2-m)^\nu
\nn\\ && \times
\tilde Z_2^\nu((m^2+s_1^2)^{1/2},(m^2+s_2^2)^{1/2}).
\ee
These simplified expressions for the unquenched resolvents have not been presented previously.

\subsection{Density of Real Eigenvalues}

The Wilson Dirac operator has as least $\nu$ real eigenvalues in the
physical sector of the Dirac spectrum. Because the eigenvalues of the
Wilson Dirac operator occur in complex conjugate pairs, additional real
eigenvalues may be produced when a pair collides with the real axis. Depending
on the expectation value of $\gamma_5$ the density 
of the real eigenvalues may be decomposed
as \cite{mario}
\be
\rho_{\rm real}^\nu(\lambda^W) = \rho_r^\nu(\lambda^W)+ \rho_l^\nu(\lambda^W).
\ee
The distribution of the chiralities
over real part of the spectrum of $D_W$  
\be
\rho_\chi^\nu(\lambda^W) = \left \langle \sum_{\lambda^W_k \in {\mathbb R}} 
\delta(\lambda^W + \lambda^W_k) {\rm sign}(\langle k| \gamma_5 | k \rangle)
\right \rangle_\nu =\rho_l^\nu(\lambda^W)- \rho_r^\nu(\lambda^W),
\ee
was derived 
from the chiral Lagrangian (\ref{zgen}) both quenched \cite{DSV,ADSV}
and with dynamical quarks \cite{ADSVNf1,SV11}.
We have that
\be
\rho_\chi^\nu(\lambda^W )\le \rho_{\rm real}^\nu (\lambda^W) 
\quad {\rm and} \quad
\int d\lambda^W \rho_\chi^\nu(\lambda^W) =\left\langle \sum_{\lambda^W_k \in {\mathbb R}} 
 {\rm sign}(\langle k| \gamma_5 | k \rangle)\right\rangle_\nu = \nu.
\ee
The expression for $\rho^\nu_\chi(\lambda^W)$  can again be simplified to 
 at most two-dimensional integrals.

The calculation of the density of real eigenvalues is more
complicated. However, starting from a random matrix theory for the
Wilson Dirac operator, it was possible to derive the joint eigenvalue
probability density of $D_W$ \cite{mario} with  
 the total number of real eigenvalues given by
\cite{mario,mario-proc}
\be
\int d\lambda^W \rho_{\rm real}^\nu( \lambda^W) =  \nu + O( (Va^2)^{\nu+1}).\nn
\ee

\section{ Conclusions}

 We have obtained the lattice spacing dependence of the
microscopic spectral density and all spectral 
correlators of the Hermitian Wilson
Dirac operator with degenerate dynamical quark masses.
This makes it possible to extract the low-energy constant
$W_8$ from the distribution of the smallest eigenvalues.
It should satisfy the constraint due to $\gamma_5$-Hermiticity 
that $W_8 >0$ (see also the recent discussion in \cite{HansenSharpe}). 
These results can be extended to include the two other low-energy
constants to this order. Our results also explain the lattice result that
the width of the distribution of the smallest eigenvalue
scales as $a/\sqrt V$. More results can be obtained
using a random matrix formulation of the chiral Lagrangian. Then
powerful random matrix methods make it possible to obtain all spectral
correlation functions of $D_W$ in the microscopic domain.

\noindent
{\bf Acknowledgments:} This work was supported  by U.S. DOE Grant No. 
DE-FG-88ER40388 (JV) and the {\sl Sapere Aude} program of The Danish
Council for Independent Research (KS).  Gernot Akemann, Poul Damgaard, 
Thomas Guhr and Mario Kieburg are thanked for useful discussions.


\begin{thebibliography}{99}

\bibitem{SV}
  E.~V.~Shuryak and J.~J.~M.~Verbaarschot,
  Nucl.\ Phys.\  A {\bf 560}, 306 (1993)
  [hep-th/9212088];
  J.~J.~M.~Verbaarschot and I.~Zahed,
  Phys.\ Rev.\ Lett.\  {\bf 70}, 3852 (1993)
  [hep-th/9303012];
  J.~J.~M.~Verbaarschot,
  Phys.\ Rev.\ Lett.\  {\bf 72}, 2531 (1994)
  [hep-th/9401059].

\bibitem{DOTV}  P.~H.~Damgaard, J.~C.~Osborn, D.~Toublan and J.~J.~M.~Verbaarschot,
  Nucl.\ Phys.\  B {\bf 547}, 305 (1999)
  [hep-th/9811212].
 
\bibitem{DSV}
  P.~H.~Damgaard, K.~Splittorff and J.~J.~M.~Verbaarschot,
  Phys.\ Rev.\ Lett.\  {\bf 105}, 162002 (2010).
  [arXiv:1001.2937 [hep-th]].

\bibitem{ADSV}
  G.~Akemann, P.~H.~Damgaard, K.~Splittorff, J.~J.~M.~Verbaarschot,
  Phys. Rev. D 83, 085014 (2011) [arXiv:1012.0752 [hep-lat]].


\bibitem{ADSVNf1}
  G.~Akemann, P.~H.~Damgaard, K.~Splittorff, J.~J.~M.~Verbaarschot,
  PoS {\bf LATTICE2010}, 079 (2010).
  [arXiv:1011.5121 [hep-lat]].

\bibitem{SV11}
  K.~Splittorff, J.~J.~M.~Verbaarschot,
  Phys.\ Rev.\  {\bf D84}, 065031 (2011).
  [arXiv:1105.6229 [hep-lat]].



\bibitem{heller}
  P.~H.~Damgaard, U.~M.~Heller, K.~Splittorff,
    [arXiv:1110.2851 [hep-lat]].
\bibitem{deuzeman}
  A.~Deuzeman, U.~Wenger, J.~Wuilloud,
  [arXiv:1110.4002 [hep-lat]].

\bibitem{Osborn} 
  J.~C.~Osborn,
  Phys.\ Rev.\  {\bf D83}, 034505 (2011).
  [arXiv:1012.4837 [hep-lat]].

 
\bibitem{Luscher}
  L.~Del Debbio, L.~Giusti, M.~L\"uscher, R.~Petronzio and N.~Tantalo,
  JHEP {\bf 0602}, 011 (2006)
  [hep-lat/0512021];
  JHEP {\bf 0702}, 056 (2007)
  [hep-lat/0610059].


\bibitem{Golterman}
  M.~Golterman, S.~R.~Sharpe, R.~L.~Singleton, Jr.,
  Phys.\ Rev.\  {\bf D71}, 094503 (2005).
  [hep-lat/0501015].
 
\bibitem{Sharpe}
  S.~R.~Sharpe,
  Phys.\ Rev.\  D {\bf 74}, 014512 (2006)
  [arXiv:hep-lat/0606002].
\bibitem{NS}
  S.~Necco, A.~Shindler,
  [arXiv:1101.1778 [hep-lat]].


\bibitem{AOSV}
  G.~Akemann, J.~C.~Osborn, K.~Splittorff, J.~J.~M.~Verbaarschot,
  Nucl.\ Phys.\  {\bf B712}, 287-324 (2005).
  [hep-th/0411030].


\bibitem{Guhr}
  T.~Guhr,
  Annals Phys.\  {\bf 250}, 145-192 (1996).


\bibitem{Kieburg}
  M.~Kieburg,
  [arXiv:1109.5109 [math-ph]].

\bibitem{mario}
  M.~Kieburg, J.~J.~M.~Verbaarschot, S.~Zafeiropoulos,
  [arXiv:1109.0656 [hep-lat]].

\bibitem{nagao}
  G.~Akemann, T.~Nagao,
  JHEP {\bf 1110}, 060 (2011).
  [arXiv:1108.3035 [math-ph]].


\bibitem{SharpeSingleton}
  S.~R.~Sharpe and R.~L.~Singleton,
  Phys.\ Rev.\  D {\bf 58}, 074501 (1998)
  [hep-lat/9804028].

\bibitem{RS} 
G.~Rupak and N.~Shoresh,
Phys.\ Rev. \ {\bf 66}, 054503 (2002), [arXiv:hep-lat/0201019].


\bibitem{BRS}
O.~B\"ar, G.~Rupak and N.~Shoresh,
Phys.\ Rev. \ D {\bf 70}, 034508 (2004), [arXiv:hep-lat/0306021].

\bibitem{Itoh}
  S.~Itoh, Y.~Iwasaki and T.~Yoshie,
  Phys.\ Rev.\  D {\bf 36}, 527 (1987).


\bibitem{Heller}
  K.~M.~Bitar, U.~M.~Heller and R.~Narayanan,
  Phys.\ Lett.\  B {\bf 418}, 167 (1998).
[arXiv:hep-th/9710052].


\bibitem{arXiv:0811.0409} 
  S.~R.~Sharpe,
  Phys.\ Rev.\ D\ {\bf 79}, 054503  (2009)
  [arXiv:0811.0409 [hep-lat]].

\bibitem{Aokiclassic}
S.~Aoki,
  Phys.\ Rev.\  D {\bf 30} (1984) 2653.

\bibitem{arXiv:1110.5744} 
  R.~N.~Larsen,
  arXiv:1110.5744 [hep-th].

\bibitem{mario-proc}
  M.~Kieburg, J.~J.~M.~Verbaarschot, S.~Zafeiropoulos,
   [arXiv:1110.2690 [hep-lat]].



\bibitem{HansenSharpe}
  M.~T.~Hansen, S.~R.~Sharpe,
    [arXiv:1111.2404 [hep-lat]].


\end{thebibliography}
\end{document}